\documentclass[12pt]{article}
\usepackage{amsmath}
\usepackage{amssymb}

\begin{document}
\title{\large{\bf Phase transitions in a frustrated Ising antiferromagnet on \\ a square lattice}}
\author{ {\bf A. Bob\'ak}$^1\,$
\footnote{Corresponding author. Fax: +421-55-2342524.
\protect\newline
\hspace*{0.3cm} \it {E-mail address:} andrej.bobak@upjs.sk
(A. Bob\'ak).},
\hspace*{0.5mm}
{\bf T. Lu\v{c}ivjansk\'y}$^{1,2}\,$, {\bf M. Borovsk\'y}$^1\,$, {\bf M. \v{Z}ukovi\v{c}}$^1\,$,\\
\\
\normalsize $^1\,$Department of Theoretical Physics and Astrophysics,
Faculty of Science,\\
\normalsize P. J. \v{S}af\'{a}rik University, Park Angelinum 9, 041 54 Ko\v{s}ice, Slovak Republic\\
\normalsize $^2\,$Fakult\"at f\"ur Physik, Universit\"at Duisburg-Essen, D-47048 Duisburg, Germany\\
}

\date{}
\maketitle
\vspace*{1cm}

{\bf Abstract}

\hspace*{0.5cm}The phase transitions occurring in the frustrated Ising square antiferromagnet with first- ($J_1 < 0$) and second- ($J_2 < 0$) nearest-neighbor interactions are studied within the framework of the effective-field theory with correlations based on the different cluster sizes and for a wide range of $R = J_2/|J_1|$. Despite the simplicity of the model, it has proved difficult to precisely determine the order of the phase transitions. In contrast to the previous effective-field study, we have found a first-order transition line in the region close to $R = -0.5$ not only between the superantiferromagnetic and paramagnetic ($R<-0.5$) but also between antiferromagnetic and paramagnetic ($R>-0.5$) phases.  

\vspace*{0.5cm}
{PACS number(s)}: 05.70.-a, 75.10.Hk

\vspace*{0.25cm}

\newpage 
{\bf 1. Introduction} \\

\hspace*{0.5cm}The Ising square lattice with nearest-neighbor (nn) interactions $(J_1)$ is a rare instance of the exactly soluble model $[1, 2]$ which shows a phase transition. However, adding next-near-neighbor (nnn) interactions $(J_2)$ or a magnetic field (or both), makes the model no longer exactly soluble and only some approximate solutions are possible to attack this more general problem. The behavior of this model is of interest, not only because of the existing theoretical questions, but also because the quasi-two-dimensional anisotropic antiferromagnet like K$_2$CoF$_4$ compound $[3]$ can be represented by Ising model at least within a fair approximation. Since the existence of nnn interactions and the magnetic field may give rise to other phases with different types of phase transitions and multicritical points $[4]$, we restrict our discussions of this model to the antiferromagnetic system with zero magnetic field. In this case a nnn antiferromagnetic interaction represents the simplest way to incorporate frustration in the square bipartite lattice. The Hamiltonian of the model with competing antiferromagnetic interactions between nn $(J_1 < 0)$ and nnn $(J_2 < 0)$ spins is given by      
\begin{equation}
\label{ham}
H = -J_1\sum_{\langle i,j\rangle}s_i s_j - J_2\sum_{\langle i,i_2\rangle}s_i s_{i_2},
\end{equation}
where $s_i = \pm 1$ and the first summation is carried out only over nn pairs of spins and the sums in the second term run over nnn pairs of spins.  We note that in the zero magnetic field under appropriate transformations, the antiferromagnetic and the ferromagnetic states are equivalent. Therefore, the results are independ of the sign of $J_1$ and may be compared with those obtained earlier for the ferromagnetic case $(J_1 > 0)$. \\
\hspace*{0.5cm}The ground state of (\ref{ham}) is well known. Namely, for the value of the frustration parameter $R\equiv J_2/|J_1| > -0.5$ the ground state is the N\'eel antiferromagnet (AF) and in the case of $R < -0.5$ the system orders in alternate ferromagnetic rows (columns) of opposite oriented spins (superantiferromagnet) $[5]$. The critical point separating these two phases lies at $R = -0.5$, where the transition temperature is suppressed to zero. At this point the ground-state is highly degenerate, a phenomenon which is responsible for the vanishing of long-range order at finite temperatures, and the system behaves like a one-dimensional Ising model $[6-9]$.  \\
\hspace*{0.5cm}Even though the investigations of the frustrated $J_1-J_2$ Ising model defined by the Hamiltonian (\ref{ham}) have a long history, its finite-temperature phase transition remains controversial until now. Indeed, the early numerical and analytic approaches supported the idea that the transition is always continuous for $R < -0.5$, but with critical exponents that vary with $R$ (weak universality) $[4, 6, 10, 11]$. However, some approximate studies $[8, 9, 12]$ and recent Monte Carlo method $[13,14]$ have found a first-order transition for $ -1.1 \lesssim R < -0.5$. On the other hand, very recent cluster mean-field calculation $[15]$ with cluster of size 4x4 also gives change the antiferromagnetic-paramagnetic transition to first-order in a narrow region of $R > -0.5$. Unfortunately, there are no clear indications from previous Monte Carlo simulations of the transition being first-order in this regime. Therefore, it may be worthwhile to examine this issue more systematically by using different size of the clusters in order to carefully analyze the effect of the cluster size on the occurrence of first-order transitions. To this end we study the $J_1-J_2$ model in its parameter space using an effective-field theory based on two-, four-, six-, and nine-spin clusters. This approach is based on the differential operator technique introduced into exact Ising spin identities and has been successfully applied to a variety of spin-$\frac {1}{2}$ and higher spin problems (for a review see, e.g., Ref. $[16])$ including a geometrically frustrated triangular lattice Ising AF $[17-19]$. It is important that the present effective-field theory allows us to treat large clusters in a simpler and more efficient computational manner. Of course, the effective-field calculation cannot give any information on the true critical exponents, although the second-neighbor interactions should in general lead to critical exponents which differ from those of the nearest-neighbor square Ising lattice $[20]$.      \\
         

{\bf 2. Theory} \\
\hspace*{0.5cm}The starting point for the statistics of an Ising spin system, which is much more useful for our purpose, is the following generalized Callen-Suzuki $[21, 22]$ exact identity
\begin{equation}
\label{identitygeneral}
{\langle O_{\{{n}\}}\rangle} = {\Bigg\langle \frac{\rm{Tr_{\{n\}}}[\it O_{\{{n}\}} \exp(-\beta {\it H_{\rm \it \{n\}}})]}{\rm{Tr_{\{n\}}}[\exp(-\beta {\it H_{\rm \it \{n\}}})]}\Bigg \rangle}, 
\end{equation}  
where the partial trace $\rm{Tr_{\{n\}}}$ is to be taken over the set $\{{n\}}$ of spin variables specified by the cluster spin Hamiltonian $\it H_{\rm \it\{n\}}$. Here $\it O_{\{n\}}$ denotes any arbitrary spin function including the set of all $\{{n\}}$ spin variables (finite cluster) and $\langle \cdots \rangle$ denotes the usual thermal average.\\
\hspace*{0.5cm}In order to take into account effects of frustration within the present effective-field theory, it is necessary to consider at least a two-spin cluster. In this approach, we select two nn spins, labeled $i$ and $j$, which interact with other nn and nnn spins from the neighborhood $[23]$. Hence, the multi-spin Hamiltonian $\it H_{\rm \it\{n\}}$ for the AF two-spin cluster ($n=2$) on the square lattice (Fig. 1(a)) is given by
\begin{equation}
\label{hampartaf}
{\it H_{\rm \it \{ij\}}}^{AF} = -J_1 s_i^As_j^B - s_i^A h_i^{AF} - s_j^B h_j^{AF}, 
\end{equation}
with 
\begin{equation}
\label{fields}
h_i^{AF} = J_1\sum_{i_1=1}^{3}s_{i_1}^B + J_2 \sum_{i_2=1}^{4}s_{i_2}^A, \quad \quad  h_j^{AF} = J_1\sum_{j_1=1}^{3}s_{j_1}^A + J_2 \sum_{j_2=1}^{4}s_{j_2}^B,
\end{equation}  
where $s_i^A$ is a spin variable on sublattice $A$ and $s_j^B$ is a spin variable on sublattice $B$, the superscript $\rm{AF}$ denotes the antiferromagnetic system, and the terms $i_1=j$ and $j_1=i$ are excluded from summations over $i_1$ and $j_1$, respectively. Now, according the exact Callen-Suzuki identity (\ref{identitygeneral}), the normalized staggered magnetization $m_{AF}$ associated with the AF two-spin cluster is given by  
\begin{equation}
\label{identmags}
m_{AF} \equiv \Bigg \langle \frac{1}{2}(s_i^A-s_j^B) \Bigg \rangle = {\Bigg\langle e^{h_i^{AF}D_x + h_j^{AF}D_y}\Bigg \rangle}f_{AF}(x,y)\Big|_{x=0,y=0}, 
\end{equation}
where $D_\mu = \partial/\partial \mu$ ($\mu = x, y$) are the differential operators and function $f_{AF}(x, y)$ is defined by 
\begin{equation}
\label{func}
f_{AF} (x, y) = \frac{\sinh\beta(x-y)}{\cosh\beta(x-y) + e^{2\beta J_1}\cosh\beta(x+y)}. 
\end{equation} 
\hspace*{0.5cm}At this point one should notice that the neighborhood of the sites $i$ and $j$ of the two-spin cluster for the $J_1-J_2$ model on a square lattice contain a set of common spins, namely the spins at the sites labeled by $(i_1, j_2)$ or $(j_1, i_2)$ in Fig. 1(a). These spins are frustrated directly within two-spin cluster theory, which is not the case of the one-spin cluster approximation. Now, taking this into account and assuming the statistical independence of lattice sites, Eq. (\ref{identmags}) may be rewritten as   
\begin{eqnarray}
\label{magaf}
m_{AF} &=& \Big[A_x(1) A_y(2) + B_x(1) B_y(2) + m_B \Big(A_x(1)B_y(2)+A_y(2)B_x(1)\Big)\Big]^2 \nonumber \\
& & \times \Big[A_y(1) A_x(2) + B_y(1) B_x(2) + m_A \Big(A_y(1)B_x(2)+A_x(2)B_y(1)\Big)\Big]^2 \nonumber \\
& & \times \Big(A_x(1) + m_B B_x(1)\Big)\Big(A_y(1) + m_A B_y(1)\Big) \nonumber \\
& & \times \Big[\Big(A_x(2) + m_A B_x(2)\Big)\Big(A_y(2) + m_B B_y(2)\Big)\Big]^2f_{AF} (x,y)\Big|_{x=0,y=0}, 
\end{eqnarray}
where $A_\mu(\nu) = \cosh(J_\nu D_\mu)$, $B_\mu(\nu) = \sinh(J_\nu D_\mu)$ ($\nu = 1, 2$), and $m_\alpha = \langle s_g^\alpha \rangle$ ($\alpha =~A$,~$B$) are the sublattice magnetizations per site. This equation of state is quite superior to that obtained from the standard mean-field theory, since in the present framework relations like $(s_g^\alpha)^{2r} = 1$ and $(s_g^\alpha)^{2r+1} = s_g^\alpha$, for all $r$, are taken exactly into account through the van der Waerden identity, $\exp(as_g^\alpha) = \cosh(a) + s_g^\alpha \sinh (a)$, while in the usual molecular-field theory all the self- and multi-spin correlations are neglected. Since for the AF $J_1-J_2$ model at zero magnetic field we have $m_{AF} = m_A = -m_B$, Eq. (\ref{magaf}) can be finally recast in the form 
\begin{equation}
\label{magm_{AF}}
m_{AF} = \sum_{{n}=0}^{4} K_{2n+1}^{AF} m_{AF}^{2n+1}, 
\end{equation}
where the coefficients $K_{2n+1}^{AF}$, which depend on $T$ and $R$, can easily be calculated within the symbolic programming by using the mathematical relation $\exp (\lambda D_x + \gamma D_y) f_{AF}(x, y) = f_{AF}(x+\lambda, y+\gamma)$. Because the final expressions for these coefficients are lengthy, their explicit form is omitted. We also note that in obtaining Eq.~(\ref{magm_{AF}}) we have made use of the fact that $f_{AF}(x, y) = - f_{AF}(-x, -y)$ and therefore only odd differential operator functions give nonzero contributions.\\   
\hspace*{0.5cm}In order to determine the phase diagram of the AF $J_1-J_2$ model, we should solve Eq. (\ref{magm_{AF}}) for a given value of the frustration parameter $R$ and look for the temperature at which the magnetization (order parameter) $m_{AF}$ goes to zero. However, for some values of $R$, the order parameter goes to zero discontinuously, i.~e., the transition becomes first order. To analyze first-order transitions, one needs to calculate the free energy for the AF and paramagnetic (P) phases and to find a point of intersection. Because the expression for the free energy in this effective-field theory does not exist, it will be extrapolated with the help of the relation for the equilibrium value of the order parameter (\ref{magm_{AF}}) as follows $[24]$:
\begin{equation}
\label{freeenergy}
F^{AF}(T, R, m) = F_0(T, R) + \frac{1}{2}\Big(1-\sum_{{n}=0}^{4} \frac{K_{2n+1}^{AF}}{n+1}{m^{2n}}\Big)m^2, 
\end{equation}
where $F_0(T,R)$ is the free energy of the disordered (paramagnetic) phase and $m$ is the order parameter which takes the value $m_{AF}$ at thermodynamic equilibrium. We note that relation (\ref{freeenergy}) corresponds to a Landau free energy expansion in the order parameter truncated in the $m^{10}$ term. \\  
\hspace*{0.5cm}Thus the equilibrium staggered magnetization is the value of the order parameter which minimizes the free energy given by Eq. (\ref{freeenergy}). Using the equilibrium condition  
\begin{equation}
\label{eqcondition}
\frac{\partial F^{AF}(T, R, m)}{\partial m}\Big|_{m=m_{AF}} = 0, 
\end{equation}
we recover Eq. (\ref{magm_{AF}}) for the equilibrium staggered magnetization. Then a critical temperature and a tricritical point, at which the phase transition changes from second order to first order, are determined by the following conditions $[25]$: (i) the second-order transition line when $1-K_1^{AF} =~0$ and $K_3^{AF} < 0,$ and (ii) the tricritical point (TCP) when $1-K_1^{AF} = 0$, $K_3^{AF} = 0,$ if $K_5^{AF} < 0$. However, the first-order phase transition line is evaluated by solving simultaneously two transcendental equations, namely the equilibrium condition (\ref{eqcondition}) and the equation $F^{AF}(T, R, m) = F_0(T, R)$, which corresponds to the point of intersection of the free energies for the AF and P phases. It is worth noticing that if we use the above given relations to obtain the critical and tricritical points, the results will coincide with those obtained from Eq.~(\ref{magm_{AF}}). As has been pointed out in $[25]$, these results justify our procedure. We also note that a similar methodology of obtaining the free energy of the model within the effective-field theory, employing the two-spin cluster, has been proposed in Ref. $[26]$ had used recently to investigate phase diagram of the frustrated Heisenberg model $[27]$. \\     
\hspace*{0.5cm}On the other hand, the choice of a two-spin cluster for the superantiferromagnet (SAF) is not unambiguous. Indeed, one can chose a cluster with two spins in the vertical or horizontal direction. As we will see below, only results obtained for the latter case are consistent with the ground-state behavior of the model. The Hamiltonian for such a choice of the cluster with two spins in the horizontal direction (see the scheme in Fig. 1(b)) is given by 
\begin{equation}
\label{hampartsaf}
{\it H_{\rm \it \{ij\}}}^{SAF} = -J_1 s_i^As_j^A - s_i^A h_i^{SAF} - s_j^A h_j^{SAF}, 
\end{equation}
where the corresponding `fields' $h_i^{SAF} $ and $h_j^{SAF}$ are given by
\begin{equation}
\label{fieldssaf}  
h_i^{SAF} = J_1\Big(s^A_{i_1=1} + \sum_{i_1=2}^{3}s_{i_1}^B\Big) + J_2 \sum_{i_2=1}^{4}s_{i_2}^B, \quad h_j^{SAF} = J_1\Big(s^A_{j_1=1} + \sum_{j_1=2}^{3}s_{j_1}^B\Big) + J_2 \sum_{j_2=1}^{4}s_{j_2}^B.
\end{equation} 
Then, in the same way as for the AF, one can derive equation of state for the SAF in the form  
\begin{eqnarray}
\label{magsaf}
m_{SAF}&=& \Big[A_x(1) A_y(2) + B_x(1) B_y(2) + m_B \Big(A_x(1)B_y(2)+A_y(2)B_x(1)\Big)\Big]^2 \nonumber \\
& & \times \Big[A_y(1) A_x(2) + B_y(1) B_x(2) + m_B \Big(A_y(1)B_x(2)+A_x(2)B_y(1)\Big)\Big]^2 \nonumber \\
& & \times \Big(A_x(1) + m_A B_x(1)\Big)\Big(A_y(1) + m_A B_y(1)\Big) \nonumber \\
& & \times \Big[\Big(A_x(2) + m_B B_x(2)\Big)\Big(A_y(2) + m_B B_y(2)\Big)\Big]^2f_{SAF} (x,y)\Big|_{x=0,y=0}, 
\end{eqnarray}
where $m_{SAF} \equiv \langle \frac{1}{2}(s_i^A+s_j^A)\rangle = m_A = -m_B$ and function $f_{SAF} (x,y)$ is given by
\begin{equation}
\label{funcSAF}
f_{SAF} (x, y) = \frac{\sinh\beta(x+y)}{\cosh\beta(x+y) + e^{-2\beta J_1}\cosh\beta(x-y)}. 
\end{equation}
Further, using the same procedure as above for the AF, we derive the free energy functional to determine numerically the transition lines, including the TCP between the SAF and P phases. \\
\hspace*{0.5cm}One can continue this series of approximations to consider larger and larger clusters and as a consequence, better results are expected. Therefore, the next step for the square lattice is a four-spin cluster ($n=4$) containing the symmetry of both the AF and SAF phases. In this case, we start from Eq. (\ref{identitygeneral}) with $\it O_{\{n\}} = (s_i^A + s_k^A -s_j^B - s_l^B)$/4 for the AF and $\it O_{\{n\}} = (s_i^A + s_j^A -s_k^B - s_l^B)$/4 for the SAF. The formalism is developed in the same way as above for the two-spin cluster. For instance, one derives for the AF the equation analogous to (\ref{identmags}), which reads 
\begin{equation}
\label{identmagsfour}
m_{AF} = {\Bigg\langle e^{h_i^{AF}D_1 + h_k^{AF}D_2 + h_j^{AF}D_3 + h_l^{AF}D_4}\Bigg \rangle}f_{AF}(x_1,x_2,x_3, x_4)\Big|_{\{{x_i=0}\}}, 
\end{equation}
where each variable $x_i (i=1,2,3,4)$ of the function $f_{AF}(x_1,x_2,x_3, x_4)$ (which, for brevity, is not presented explicitly here)  is associated with a differential operator $D_i =\partial/\partial x_i (i=1,2,3,4)$. The corresponding `fields' $h_i^{AF}$, $h_j^{AF}$, $h_k^{AF} $, and $h_l^{AF}$ (see also $[12]$) are given by  
\begin{eqnarray}
\label{fieldsaffour}
h_i^{AF}&=&J_1\sum_{i_1=1}^{2}s_{i_1}^B + J_2 \sum_{i_2=1}^{3}s_{i_2}^A, \quad h_j^{AF} = J_1\sum_{j_1=1}^{2}s_{j_1}^A + J_2 \sum_{j_2=1}^{3}s_{j_2}^B, \nonumber \\
h_k^{AF}&=&J_1\sum_{k_1=1}^{2}s_{k_1}^B + J_2 \sum_{k_2=1}^{3}s_{k_2}^A, \quad h_l^{AF} = J_1\sum_{l_1=1}^{2}s_{l_1}^A + J_2 \sum_{l_2=1}^{3}s_{l_2}^B.
\end{eqnarray}
Then, using of the same procedure as for the two-spin cluster, one determines the second- and first-order transition lines including the TCP between the AF and P phases. Finally, to get a more convincing evidence of the existence of a TCP in the phase diagram, we also consider six- and nine-spin clusters. In this case, the clusters consist of two or four square plaquettes, respectively, and contain more information about the lattice topology. However, the analytical calculations for such large clusters would have been very lengthy and tedious, therefore, the results were obtained in a completely numerical way within the symbolic programming by using Mathematica software package. It should be noted here that the calculation times for the large clusters become rather long even using the symbolic programming. Therefore, the highest approximation used to study the frustrated Ising square lattice is the one based on the nine-spin cluster.  \\
            
{\bf 3. Results} \\

\hspace*{0.5cm}The numerical results for the critical temperature $k_BT_N/|J_1|$ versus $R$ for various cluster sizes are shown in Figs. $2, 3$, and $5$. In these figures the solid lines indicate the second-order phase transitions, while the dashed lines represent the first-order ones. The black circles denote the positions of TCPs at which the phase transitions change from second to first order ones. \\
\hspace*{0.5cm}First, from all the figures it can be seen that the first-order transition temperatures $T_N$ between the AF and P phases as functions of the frustration parameter $R$, for the considered cluster sizes, approaches zero when $R = -0.5$, in agreement with the ground-state result. For the nonfrustrated model ($R = 0$), we find that values of $k_BT_N/|J_1|$ are $3.0250, 2.9197, 2.8759$, and $2.8185$ for the two-, four-, six-, and nine-spin clusters, respectively, which shows a relatively slow convergence to the exact value of $k_BT_N/|J_1| = 2.2692$ with increasing cluster size. Further, it is seen that the critical temperature decreases gradually to the TCP, when the frustration parameter approaches some negative value $R_t$. Our estimates for the coordinates of the TCP ($k_BT_t/|J_1|; R_t$) are $(1.3720; -0.2973), (1.3066; -0.3070), (1.2329; -0.3240)$, and $(1.1894;$ $-0.3340)$ for the two-, four-, six-, and nine-spin cluster approximations, respectively. Thus, within the present approach the tricritical temperature decreases with increasing cluster size. However, the relative decrease of the tricritical temperature becomes gradually smaller with the cluster size. For instance, the tricritical temperature for the four-spin cluster is about $5.00\%$ lower than that for the two-spin cluster but for the nine-spin cluster the derease is only $3.66\%$ in comparison with the six-spin cluster. These results indicate that there could be a narrow region of first-order transitions for $R > -0.5$ and that the cluster-size has a considerable effect on the existence and location of the TCP at which the phase transition between the AF and P phases changes from second order to first one. It should be noted here that the tricritical behavior in this region has also been predicted by a cluster mean-field approach based on the $4$x$4$ cluster but not for the $2$x$2$ cluster $[15]$. Such behavior of the model for $R > -0.5$ is also supported by the fact that at $R = -0.5$ there is, as discussed above, a first-order transition at $T = 0$ between the AF and SAF phases. We believe that the first-order behavior of the model in the region $R > -0.5$ cannot be completely ruled out and should be examined close to $R = -0.5$ not only by various approximate theories but also by more reliable techniques, such as e.g. Monte Carlo approach, to definitely confirm or to rule out this scenario. Note that this model has also been investigated by the effective-field theory based on the four-spin cluster in $[12]$. Unfortunately, some of the numerical results obtained in that work are defective, and the authors overlooked the existence of the TCP on the transition line between the ferromagnetic (or antiferromagnetic) and paramagnetic phases. It is worth mentioning that we also checked numerically that the obtained phase diagrams are independent of the sign of $J_1$. Of course, for $J_1 > 0$ (ferromagnetic case), the order parameter $m_F$ must be defined as the direct (not staggered) magnetization. \\         
\hspace*{0.5cm}On the other hand, it is generally accepted now that the transition line between the SAF and P phases is a second order for the full range of $R \lesssim -1.1$ and a first order for $ -1.1 \lesssim R < -0.5$. Such behavior is confirmed not only by approximate methods $[8, 9, 12]$, but also by recent Monte Carlo studies $[13-15, 28]$. We note that the region of $-1.1 \lesssim R < -0.67$ turns out to be very difficult to study by numerical means, nevertheless, the results of the recent study $[28]$ indicate only pseudo-first order behavior in this region but the true nature should be second order. Our effective-field results obtained for various cluster sizes support this picture. However, as seen from Fig. $2$, the transition line as well as the coordinates of the TCP obtained within the effective-field theory based on the two-spin cluster for $R < -0.5$ depend on the selection of the two neighboring spins. Namely, if we chose a cluster with two spins in the horizontal direction (see the scheme in Fig. 1(b)), the coordinates of the TCP are $(2.7178; -1.1098)$, while if consider a cluster with two spins in the vertical direction (the scheme similar to that of Fig. 1(b) is not shown here), the coordinates of the TCP are $(2.2594; -0.9991)$. Moreover, the first-order transition for the latter case terminates at $R = -0.6168$ (see the inset of Fig.~$2$), which is inconsistent with the ground-state behavior of the model. Therefore, the effective-field theory based on the cluster with two spins chosen in the vertical direction is not appropriate for the SAF state due to its incompatibility with the symmetry of the SAF state. \\
\hspace*{0.5cm}It is evident, that this lack of symmetry for the SAF state is absent in the four- or nine-spin clusters. Consequently, one can find that the corresponding first-order phase transitions terminate at $R = -0.5$, as expected from the ground-state arguments. Since the phase diagrams for these clusters are qualitatively the same, in Fig. $3$ we show only the phase diagram for the nine-spin cluster. Note also that the TCPs for the four- and nine-spin clusters are located uniqly and their coordinates are $(2.4620;-1.0387)$ and $(2.2565; -0.9748)$, respectively. The values $R_t= -1.0387$ and $R_t =-0.9748$ may be compared to those of the cluster-variation method ($R_t \approx -1.144$) $[8, 9]$, Monte Carlo study ($R_t \approx -0.9$) $[14]$, and recent Monte Carlo approach ($R_t \approx -0.67$) $[15, 28]$. On the other hand, for the effective-field theory based on the six-spin cluster there are two possibilities how to select the cluster for the SAF, namely in the horizontal or in the vertical direction (see Fig.~$4$). It is clear now that the clusters consist of different number of spins on sublattices $A$ and $B$. Consequently, the coordinates of the TCP are $(2.4018;-1.0111)$ or $(2.2194; -0.9666)$, respectively, and the corresponding phase diagrams are shown in Fig. $5$. But now both clusters consist of two square plaquettes, which are symmetrical with respect to the SAF state. Therefore, the phase diagrams for these clusters differ only quantitatively and end at $R = -0.5$, which is exact.  \\
\hspace*{0.5cm}In order to confirm the prediction of the first- and second-order phase transitions, let us examine temperature dependencies of the order parameters $m_{AF}$ and $m_{SAF}$ for the nine-spin cluster, when the value of $R$ is changed. As seen from Fig. 6(a), the order parameter $m_{AF}$ falls smoothly to zero when temperature increases from zero to $k_BT_N/|J_1|$, characterizing a second-order phase transition. Similarly, the $m_{AF}$ also reduced to zero continuously at the TCP (see curve labeled $-0.3340$). On the other hand, below the TCP, the stable solution of $m_{AF}$ becomes discontinuous at the first-order phase transition and this discontinuity increases with $R$ going to $-0.5$. The curves for $R = -0.35$ and $R= -0.4$ are examples of such behavior, where the first-order transition is indicated by a vertical dashed line. A qualitatively similar behavior exhibits the order parameter $m_{SAF}$ (Fig. 6(b)). \\       

{\bf 4. Conclusions}\\

\hspace*{0.5cm}In this work, we have investigated the effects of the frustration parameter $R = J_2/|J_1|$ on the phase diagram of the Ising antiferromagnet on the square lattice by the use of the effective-field theory with correlations and approximations based on different cluster sizes. For different finite-clusters approximations it is found that the phase transitions between the AF and P phases ($R > -0.5$) close to $R = -0.5$ are of first order. This behavior was confirmed by solving the Hamiltonian using two-, four-, six-, and nine-spin clusters. Our results indicate that the cluster size plays an important role for determining the location of the TCP in the frustrated square lattice. This behavior is consistent with that obtained within a variational mean-field theory $[15]$, which predicts the TCP for the $4$x$4$ cluster but not for the $2$x$2$ cluster. Of course, these are the effective-field results, therefore, further Monte Carlo simulations or more reliable calculations in this region are called for.  \\
\hspace*{0.5cm}On the other hand, we have shown that the frustrated square lattice exhibits the TCP, at which the phase transition changes from second order to first order, on the line between SAF and P phases, in agreement with previous approximate as well as Monte Carlo studies. However, within present effective-field theory, the position of the TCP for some clusters (two- and six-spin clusters) depends on the symmetry of the selected cluster. This behavior is due to a lower symmetry of the SAF phase in comparison to the AF phase.           \\

{\bf Acknowledgments} \\

\hspace*{0.5cm}This work was supported by the Scientific Grant Agency of Ministry of Education of Slovak Republic (Grant VEGA No.1/0234/12).\\

\newpage{}

\newpage
{\bf Figure captions} \\

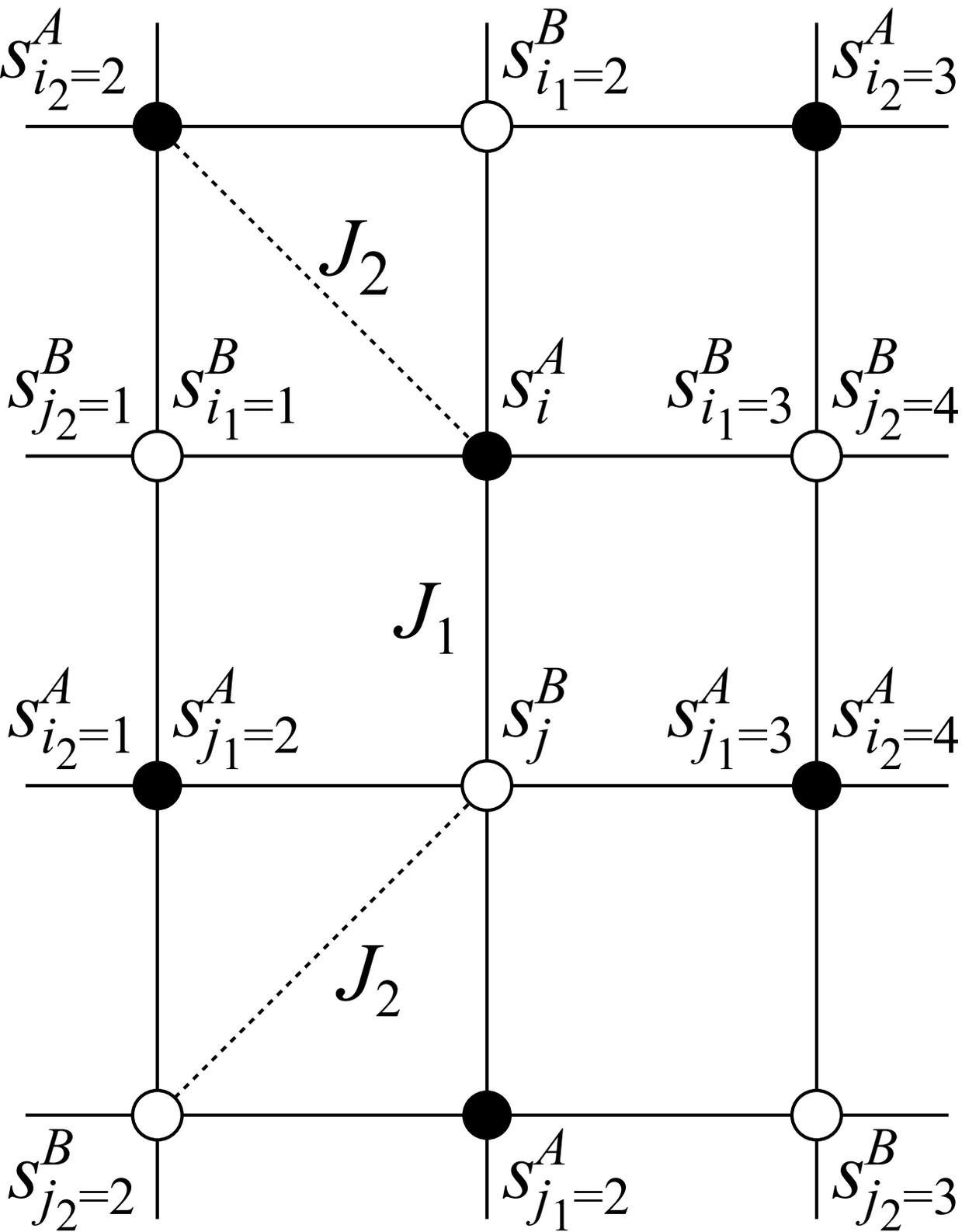
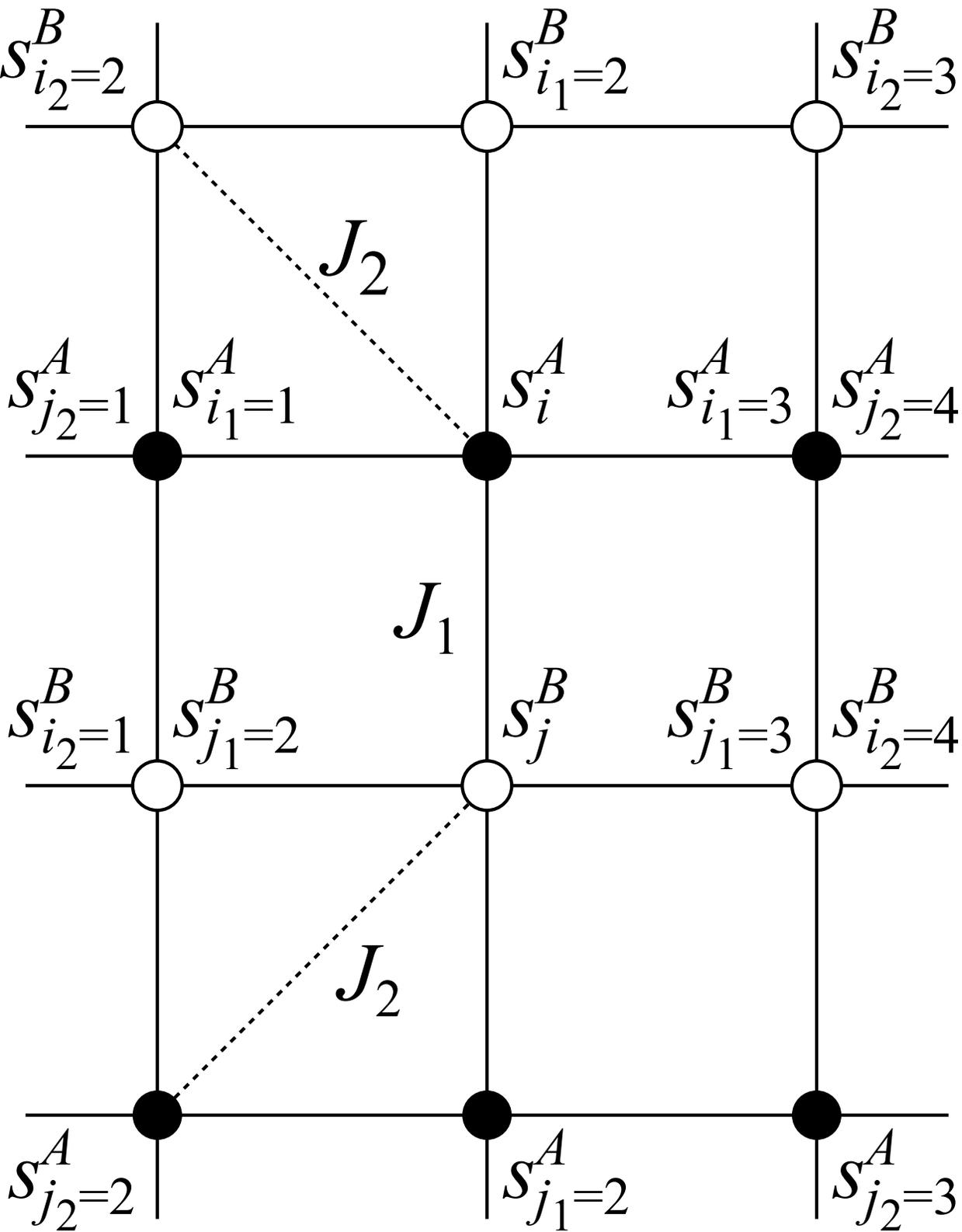
{\bf Figure 1:} Ground-state configurations of the $J_1-J_2$ Ising model on the square lattice showing (a) aniferromagnetic and (b) superantiferromagnetic states for the two-spin cluster approximation defined by the Hamiltonians (\ref{hampartaf}) and (\ref{hampartsaf}), respectively. Two sublattices are marked by black and white circles.\\

{\bf Figure 2:} Phase diagram in the coupling-temperature plane for the $J_1-J_2$ Ising model on the square lattice based on the two-spin cluster. Solid and dashed lines indicate second- and first-order transitions, respectively, while the black circles denote the position of a tricritical point. $AF$ and $SAF$ are the antiferromagnetic and superantiferromagnetic ordered phases, respectively, and $P$ is the paramagnetic phase. The inset shows phase diagram when the two-spin cluster is chosen in the vertical direction. \\

{\bf Figure 3:} Phase diagram in the coupling-temperature plane for the $J_1-J_2$ Ising model on the square lattice based on the nine-spin cluster. Solid and dashed lines indicate second- and first-order transitions, respectively, while the black circles denote the position of a tricritical point. $AF$ and $SAF$ are the antiferromagnetic and superantiferromagnetic ordered phases, respectively, and $P$ is the paramagnetic phase. \\

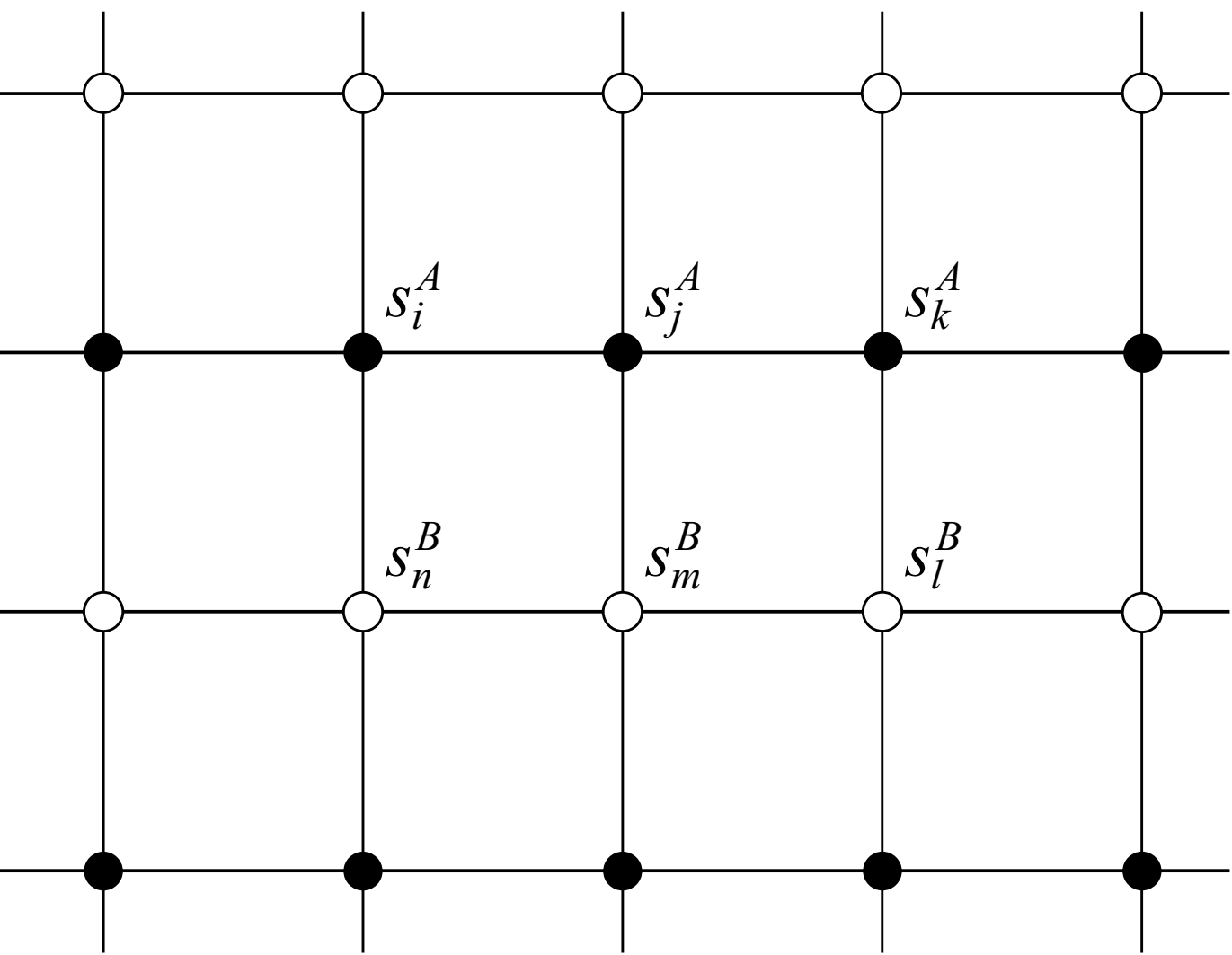
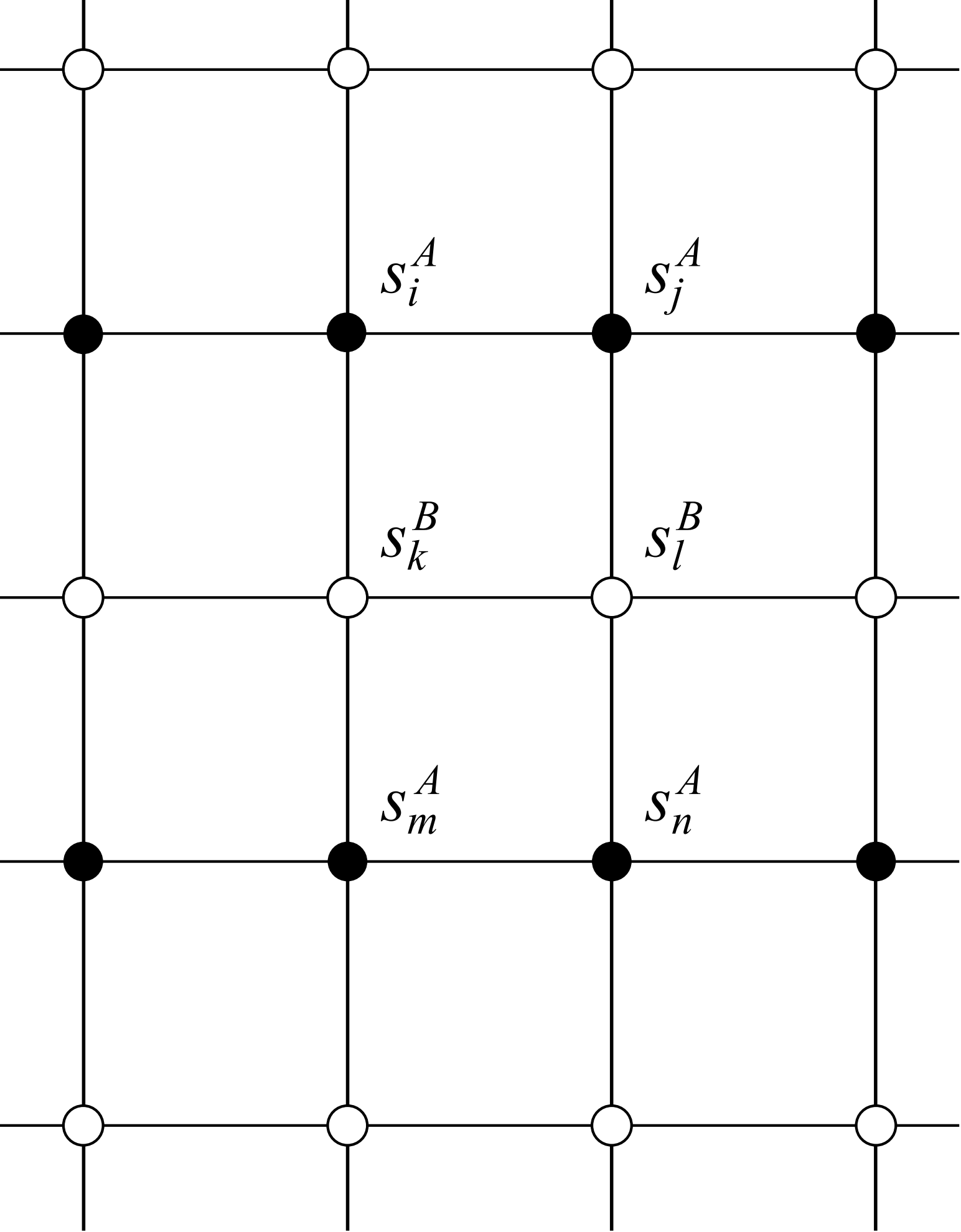
{\bf Figure 4:} Two options of the six-spin cluster with spins $s_i, s_j, s_k, s_l, s_m$, and $s_n $ for the superantiferromagnetic arrangement on the square lattice: (a) in the horizontal direction and (b) in the vertical direction. Two sublattices are marked by black and white circles. Each cluster has $14$ nn and nnn spins.\\

{\bf Figure 5:} The same as in Fig. 2, but for the six-spin cluster.\\

{\bf Figure 6:} Temperature dependences of the order parameters of $m_{AF}$ in (a) and $m_{SAF}$ in (b) for the $J_1-J_2$ Ising model on the square lattice based on the nine-spin cluster, when the frustration parameter $R$ is changed. The dashed lines indicate the first-order transitions. \\


\begin{thebibliography}{12}
\bibitem{}
L. Onsager, Phys. Rev. {\bf 65}, 117 (1944).
\bibitem{}
B. M. McCoy and T. T. Wu, {\it The two-dimensional Ising model} (Harvard University, Cambridge, Mass., 1973).
\bibitem{}
H. Ikeda and H. Hirakawa, Solid State Commun. {\bf 14}, 529 (1979).
\bibitem{}
K. Binder and D.P. Landau, Phys. Rev. B {\bf 21}, 1941 (1980).
\bibitem{}
C. Fan and F.Y. Wu, Phys. Rev. {\bf 179}, 560 (1969). 
\bibitem{}
D.P. Landau, Phys. Rev. B {\bf 21}, 1285 (1980).
\bibitem{}
M.D. Grynberg and B. Tanatar, Phys. Rev. B {\bf 45}, 2876 (1992).
\bibitem{}
J.L. Mor\'an-L\'opez, F. Aguilera-Granja, and J.M. Sanchez, Phys. Rev. B {\bf 48}, 3519 (1993). 
\bibitem{}
J.L. Mor\'an-L\'opez, F. Aguilera-Granja, and J.M. Sanchez, J. Phys.: Condens. Matter {\bf 6}, 9759 (1994).
\bibitem{}
R.H. Swendsen and S. Krinsky, Phys. Rev. Lett. {\bf 43}, 177 (1979).
\bibitem{}
J. Oitmaa, J. Phys. A {\bf 14}, 1159 (1981).
\bibitem{}
R.A. dos Anjos, J.R. Viana, and J.R. de Sousa, Phys. Lett. A {\bf 372}, 1180 (2008).
\bibitem{}
A. Kalz, A. Honecker, S. Fuchs, and T. Pruschke, Eur. Phys. J. B {\bf 65}, 533 (2008).
\bibitem{}
A. Kalz, A. Honecker, and M. Moliner, Phys. Rev. B {\bf 84}, 174407 (2011).
\bibitem{}
S. Jin, A. Sen, W. Guo, and A.W. Sandvik, Phys. Rev. B {\bf 87}, 144406 (2013). 
\bibitem{}
T. Kaneyoshi, Acta Phys. Polonica A {\bf 83}, 703 (1993).
\bibitem{}
M. \v{Z}ukovi\v{c}, M Borovsk\'y, and A. Bob\'ak, Phys. Lett. A {\bf 374}, 4260 (2010). 
\bibitem{}
M. \v{Z}ukovi\v{c}, M Borovsk\'y, and A. Bob\'ak, J. Magn. Magn. Mater. {\bf 324}, 2687 (2012).
\bibitem{}
M. Borovsk\'y, M. \v{Z}ukovi\v{c}, and A. Bob\'ak, Physica A {\bf 392}, 157 (2013).
\bibitem{}
F. W. Wu, Phys. Rev. B {\bf 4}, 2312 (1971). 
\bibitem{}
H.B. Callen, Phys. Lett. {\bf 4}, 161 (1963).
\bibitem{}
M. Suzuki, Phys. Lett. {\bf 19}, 267 (1965).
\bibitem{}
A. Bob\'ak and M. Ja\v{s}\v{c}ur, Phys. Status Solidi B {\bf 135}, K9 (1986).
\bibitem{}
P.R. Silva and F.C. S\'a Barreto, Phys. Status Solidi B {\bf 114}, 227 (1982).
\bibitem{}
A. Bob\'ak and P. Macko, J. Magn. Magn. Mater. {\bf 109}, 172 (1992).
\bibitem{}
R.A. dos Anjos, J.R. Viana, J.R. de Sousa, and J.A. Plascak, Phys. Rev. E {\bf 76}, 022103 (2007).
\bibitem{}
R.S. Lapa, G. Mendon\c{c}a, J.R. Viana, and J.R. de Sousa, J. Magn. Magn. Mater. {\bf 369}, 44 (2014).
\bibitem{}
S. Jin, A. Sen, and A.W. Sandvik, Phys. Rev. Lett. {\bf 108}, 045702 (2012).
\end{thebibliography}
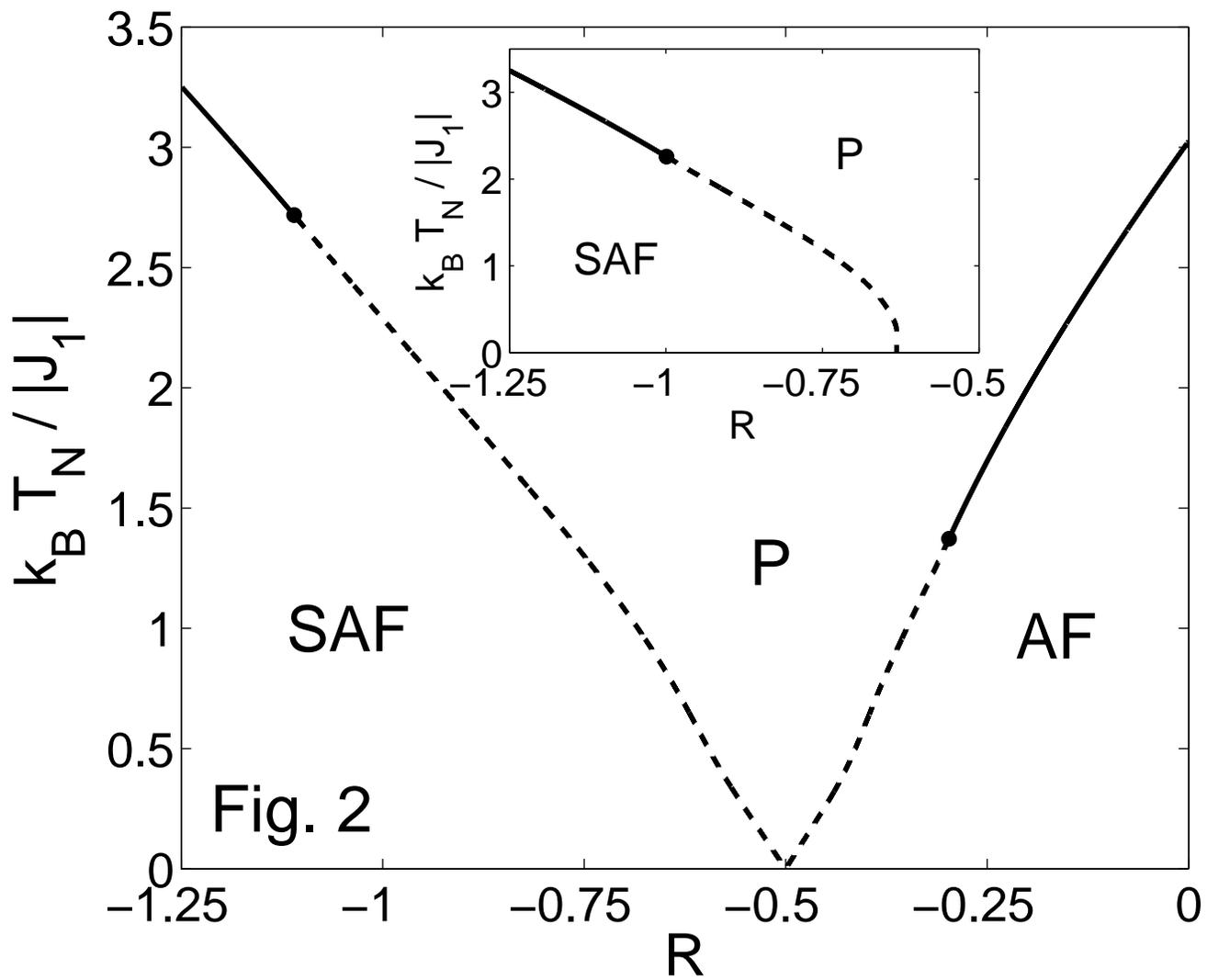
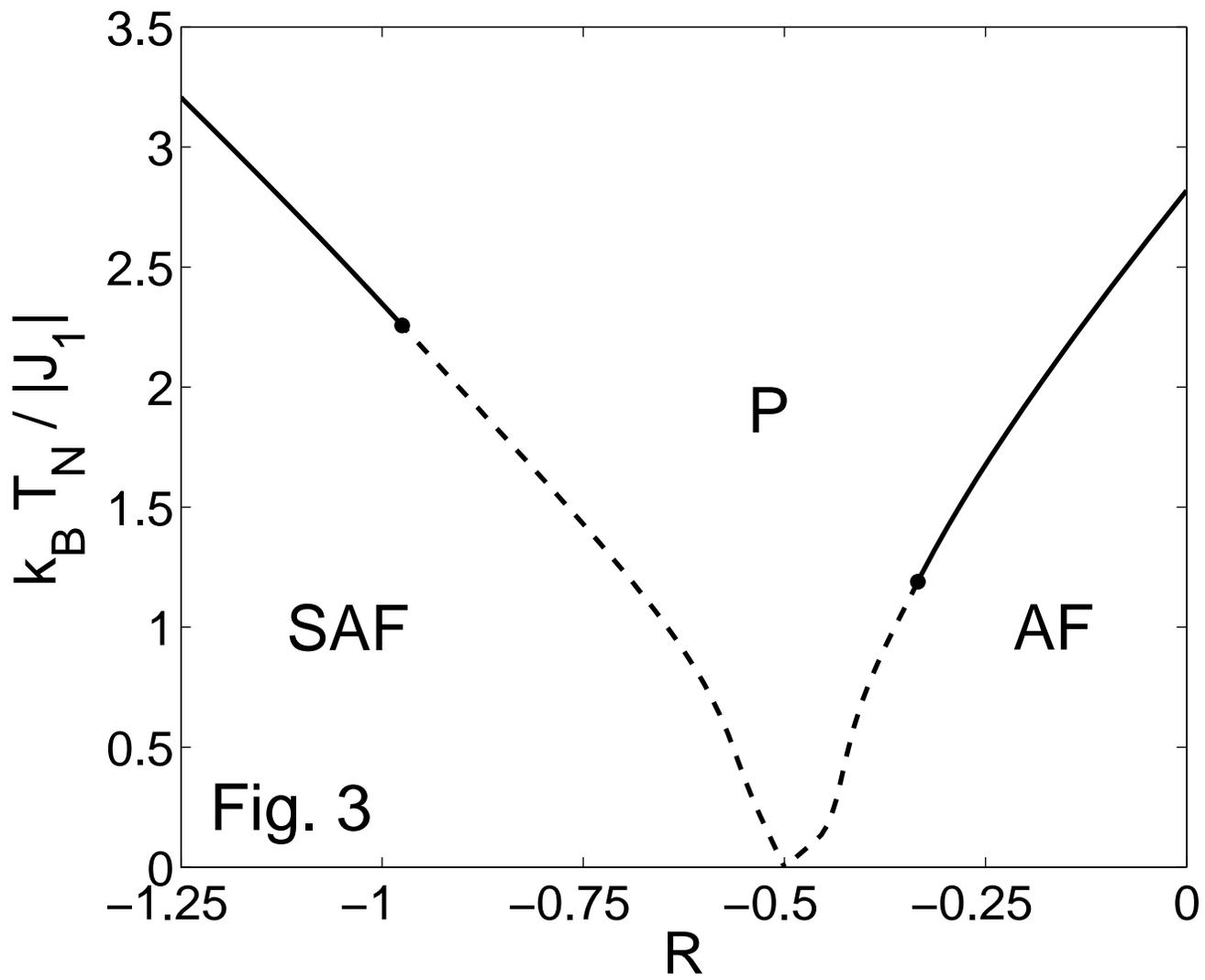
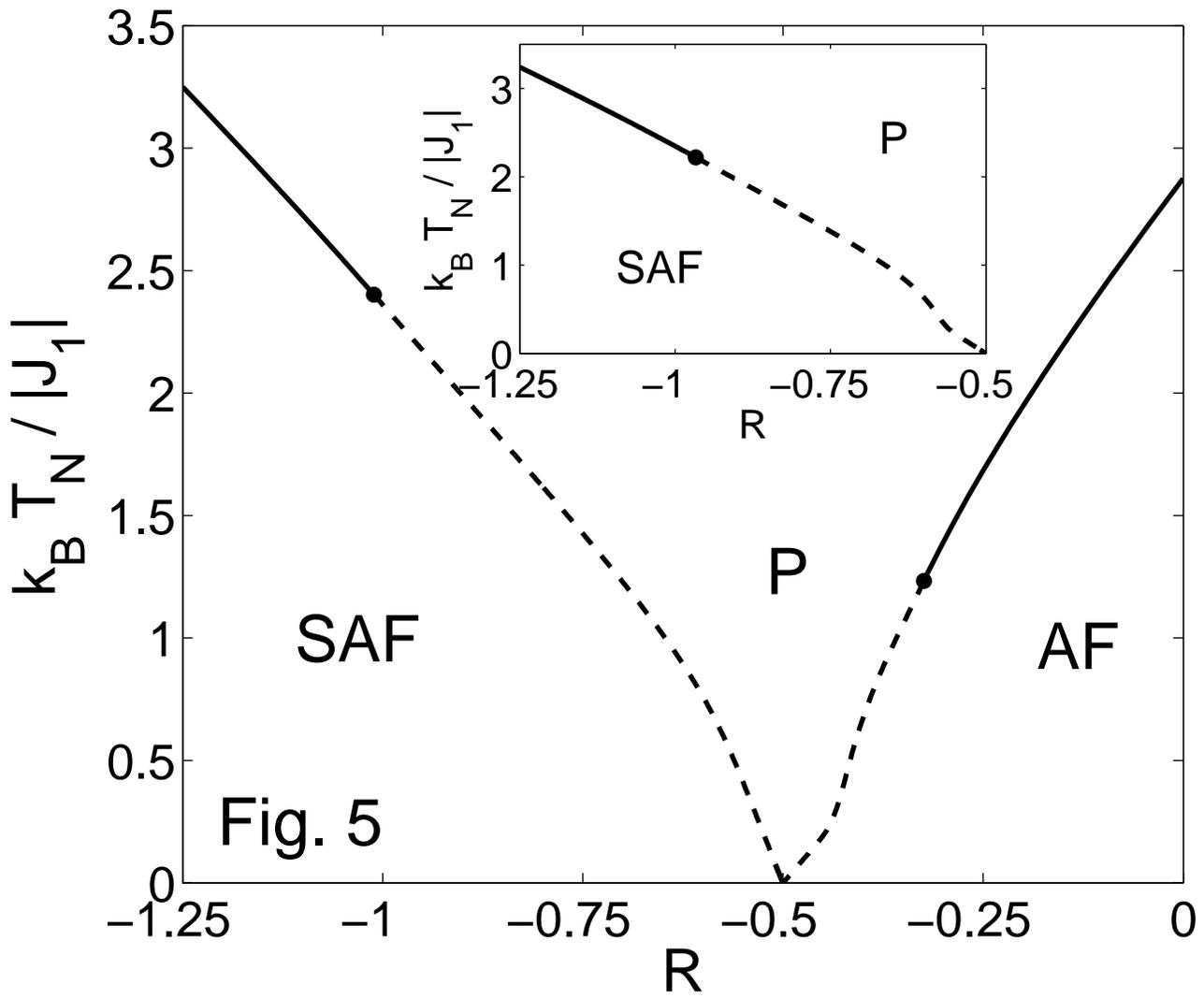
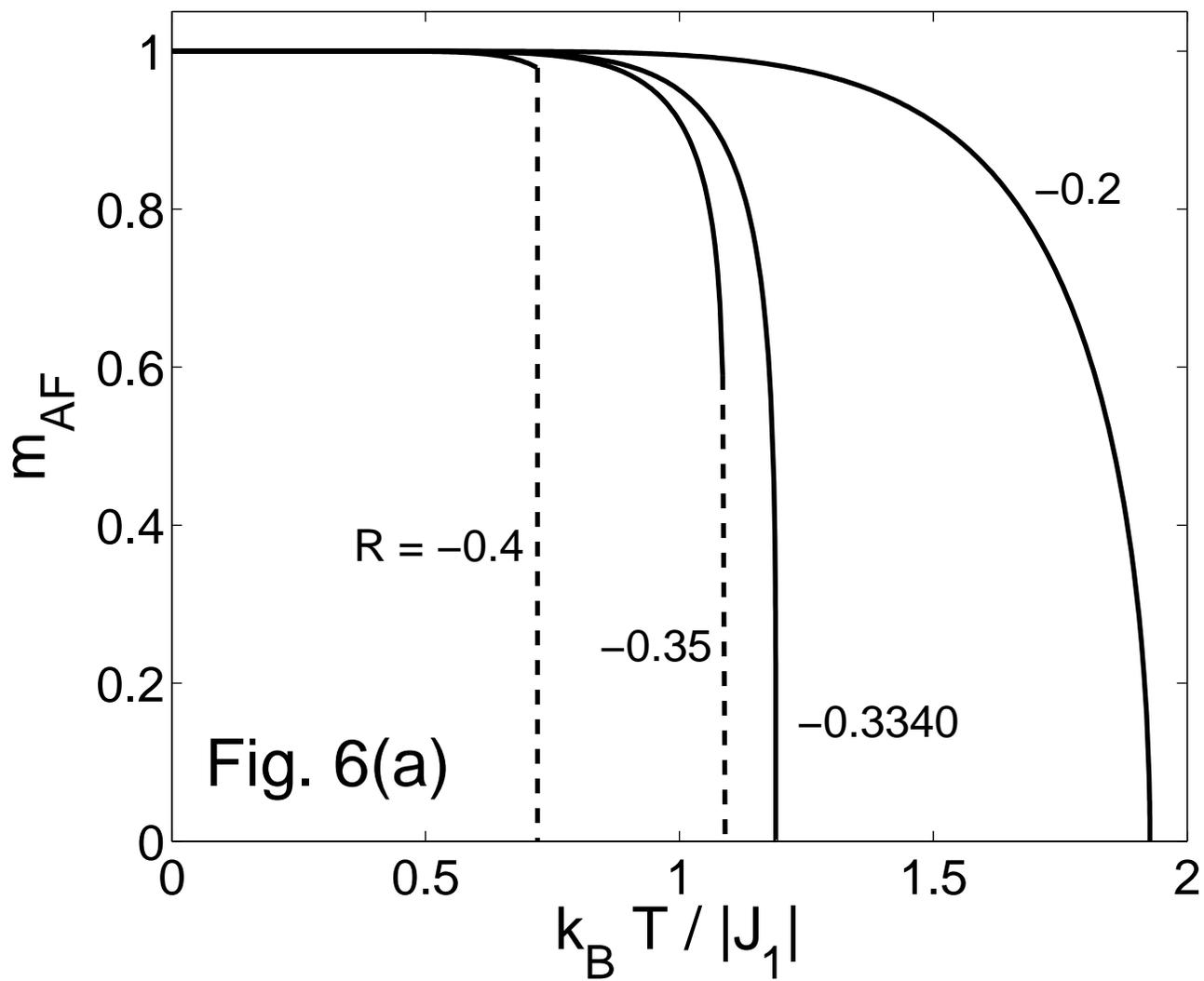
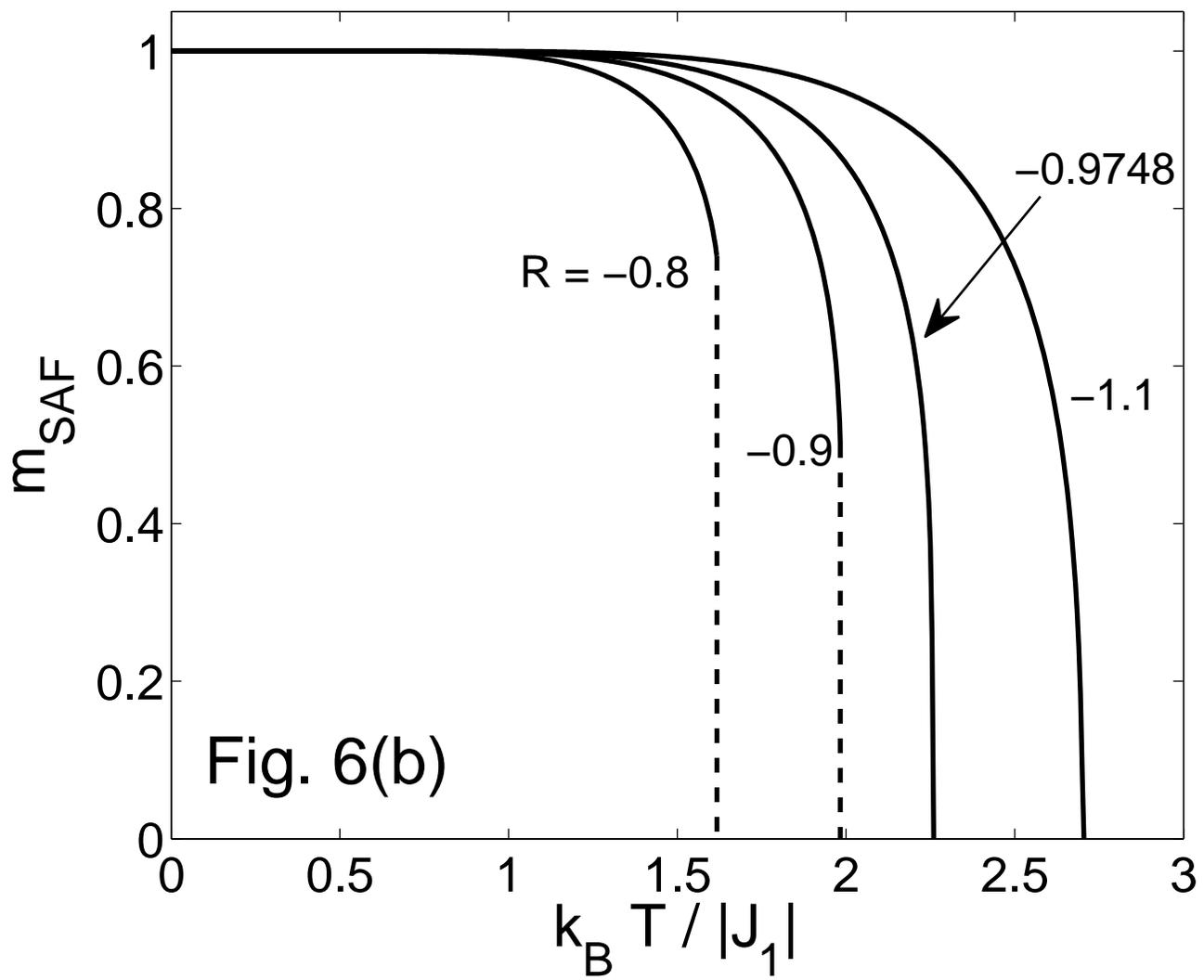
\end{document}